\documentstyle[12pt]{article}
\begin{document}
\input epsf
\begin{titlepage}
\begin{center}
\today     \hfill    WM-99-103\\
%~{} \hfill UCB-PTH-DRAFT \\

\vskip .5in

{\large \bf Gauge Unification in Nonminimal Models with Extra Dimensions}

\vskip 0.3in

Christopher D. Carone\footnote{carone@physics.wm.edu} 

\vskip 0.1in

{\em Nuclear and Particle Theory Group \\
     Department of Physics \\
     College of William and Mary \\
     Williamsburg, VA 23187-8795}
%% NOTE THAT I RESET THE equation COUNTER AT THE BEGINNING
%% OF EACH SECTION, SO THAT EQUATION NUMBERING IS OF THE
%% FORM  (Section.equation)

%\vskip 0.1in

%{{}$^2$ \em Department of Physics\\
%     University of California, Berkeley, California 94720}
        
\end{center}

\vskip .5in

\begin{abstract}
We consider gauge unification in nonminimal models with extra spacetime 
dimensions above the TeV scale.  We study the possibility that only a
a subset of the supersymmetric standard model gauge and Higgs fields live 
in the higher dimensional `bulk'.  In two of the models we present, a 
choice for bulk MSSM matter fields can be found that preserves 
approximate gauge unification.  This is true without the addition of 
any exotic matter multiplets, beyond the chiral conjugate mirror fields 
required to make the Kaluza-Klein excitations of the matter fields 
vector-like.  In a third model, only a small sector of additional matter 
fields is introduced. In each of these examples  we show that gauge 
unification can be obtained without the necessity of large string scale 
threshold corrections.  We comment briefly on the phenomenology of these 
models in the case where the compactification scale is as low as possible.
\end{abstract}
\end{titlepage}

\newpage
\renewcommand{\thepage}{\arabic{page}}
\setcounter{page}{1}
\section{Introduction} \label{sec:intro} \setcounter{equation}{0}

Recently, Dienes, Dudas and Gherghetta (DDG) \cite{ddg,dnew} have suggested
the intriguing possibility that grand unification may occur at 
intermediate or even low energy scales in models with extra spacetime
dimensions compactified on orbifolds of radii $R$.  Above the 
compactification scale $\mu_0 = 1/R$, the vacuum polarization tensors
for the gauge fields of the minimal supersymmetric standard model (MSSM)
receive finite corrections from a tower of Kaluza-Klein (KK) excitations
which contribute at one loop.  As a consequence, the gauge couplings develop
a power-law rather than logarithmic dependence on the ultraviolet cut off
of the theory, $\Lambda$ \cite{ddg,taylor}.  While this is not running in 
the conventional sense, the gauge couplings nonetheless evolve rapidly as 
a function of $\Lambda$, so that it is possible to achieve an accelerated 
unification.

In the minimal scenario proposed by DDG, all the non-chiral MSSM fields,
the two Higgs doublets and the gauge multiplets, live in a $4+\delta$
dimensional spacetime, and have an associated tower of KK excitations.
The chiral MSSM fields are assumed to lie at fixed points of the
orbifolds, and thus have no KK towers.  This is the simplest
way of avoiding the difficulties associated with giving mass
to chiral KK states.  DDG demonstrated that an approximate gauge
unification could be achieved in this scenario, at a grand
unification scale $M_{{\rm GUT}}$ that is much smaller than 
its usual value in supersymmetric theories, $2\times10^{16}$~GeV.  
However, as pointed out in Ref.~\cite{dnew,ross}, strict
comparison to the low energy data reveals that for TeV-scale
compactifications, the DDG model predicts a value for $\alpha_3(m_z)$
that is higher than the prediction in conventional unified theories,
which is already $\sim$5 standard deviations higher than the 
experimental value\footnote{Using the same two loop code and input values
described later in this paper, we find that $\alpha_3(m_z)\approx 0.1276$
in conventional supersymmetric unified theories, compared to the world
average, $0.1191\pm 0.0018$ \cite{rpp98}.}.  Assuming that the unification 
point coincides with the string scale, then a specific model of string scale 
threshold corrections is required before one can claim that unification
in the DDG model is actually achieved.

It is the purpose of this paper to point out that there are a number
of simple variations on the DDG proposal that achieve gauge unification
much more precisely than the minimal scenario described 
above.  To begin, however, let us consider the variations that are 
as successful as the minimal case.  As pointed out by DDG, one 
possibility is to allow $\eta$ generations of matter fields to experience 
extra dimensions, and to add to the theory their chiral conjugate mirror
fields, so that suitable KK mass terms may be formed.  Assuming that
the orbifold is $S^1/Z_2$, then one may take the mirror fields to be
$Z_2$ odd, so that unwanted zero modes are not present in the low-energy
theory.  Given that the KK excitations of the matter fields form complete 
SU(5) multiplets, it is perhaps not surprising that if unification is 
achieved for $\eta=0$, it will also be preserved for $\eta\neq 0$, at least 
for some range of $\mu_0$.

What is less obvious is that the KK excitations of the matter fields
may be chosen to form {\em incomplete} SU(5) multiplets, and an
approximate unification may still be preserved if only {\em some} of the 
MSSM gauge and Higgs fields experience extra dimensions.  In Section~2, we
will present three models that demonstrate (i) that it is not
necessary for all gauge groups to live in the higher dimensional
bulk in order to achieve unification, and (ii) a precise unification may 
sometimes be obtained without introducing any exotic matter, beyond the
mirror fields described above\footnote{For other possibilities, 
see Ref.~\cite{kaku}.}. In Section~3  we study these cases 
quantitatively, taking into account weak-scale threshold corrections, 
and two-loop running up to the compactification scale.  For TeV scale
compactifications \cite{ant}, we will see that all of the nonminimal 
scenarios we present in this paper unify more precisely than the 
minimal DDG scenario, and do not require large threshold corrections at 
the unification scale. In Section~4 we summarize our conclusions, and 
make some brief comments on the phenomenological implications of these 
models when the compactification scale is low.

\section{Three scenarios}

We assume $4+\delta$ spacetime dimensions, with $\delta$ dimensions
each compactified on a $Z_2$ orbifold of radius $1/\mu_0$.  The fields
that experience extra dimensions are periodic in the $\delta$ new
spacetime coordinates $y_1 \ldots y_\delta$, and are either even or
odd under $\vec{y} \rightarrow -\vec{y}$.  For example, in the case 
where $\delta=1$, these `bulk' fields have expansions of the form
\[
\Phi_+ = \sum_{n=0}^\infty \cos(\frac{n y}{R}) 
\Phi^{(n)}(x^\mu) \,\,\, ,
\]
\begin{equation}
\Phi_- = \sum_{n=1}^\infty \sin(\frac{n y}{R}) 
\Phi^{(n)}(x^\mu)
\end{equation}
where $n$ indicates the KK mode.  The only other fields in the theory
are those which live at the orbifold fixed points $y=0$ or $y=\pi R$,
and have no KK excitations.

The effect of a tower of KK states on the running of the MSSM gauge
couplings was computed by DDG, and is given in a useful approximate
form by \cite{ddg}
\begin{eqnarray}
\alpha_i^{-1}(\Lambda) & = &\alpha_i^{-1}(m_z)-\frac{b_i}{2\pi}
\ln(\frac{\Lambda}{m_z})+\frac{\tilde{b}_i}{2\pi}\ln(\frac{
\Lambda}{\mu_0})\\ &-&\frac{\tilde{b}_i X_\delta}{2\pi\delta}\left[
(\frac{\Lambda}{\mu_0})^\delta-1\right] \,\,\, .
\label{eq:drun}
\end{eqnarray}
Here the $\tilde{b}_i$ are the beta function contributions of a
single KK level, and $X_\delta$ is given by
\begin{equation}
X_\delta = \frac{2\pi^{\delta/2}}{\delta\Gamma(\delta/2)} \,\,\, .
\end{equation}
For the scenarios considered by DDG, these beta functions are 
\begin{equation}
b_i=(\frac{33}{5},1,-3) \,\,\,\,\,\,\,\,\,\, 
\tilde{b}_i=(\frac{3}{5},-3,-6)+\eta (4,4,4) \,\,\, ,
\label{eq:bmin}
\end{equation}
where $\eta$ is the number of generations of matter fields that
experience extra dimensions.  DDG observe that a sufficient condition
for gauge unification to be preserved is that the ratios
\begin{equation}
B_{ij}=\frac{\tilde{b}_i-\tilde{b}_j}{b_i-b_j}
\end{equation}
be independent of $i$ and $j$.  Thus, they point out that in
the scenario above
\begin{equation}
\frac{B_{12}}{B_{13}}=\frac{72}{77}\approx 0.94
\,\,\,\,\,\mbox{ and }\,\,\,\,\,
\frac{B_{13}}{B_{23}}=\frac{11}{12}\approx 0.92 \,\,\, .
\end{equation}
We will now show that there are a variety of other models, each 
with a different set of MSSM fields living at the orbifold 
fixed points, that lead to $B_{12}/B_{13} \approx B_{13}/B_{23}
\approx 1$.  First, notice that the $\tilde{b}_i$ of the minimal scenario 
can be decomposed into the contributions from the KK excitations of 
each MSSM field, as shown in Table~1.  An overline denotes a mirror
field required so that the given KK tower is vector-like.
\begin{table}
\begin{center}
\begin{tabular}{lll} \hline\hline
gauge & (0,-4,-6) & \\
$H_u + H_d$ & (3/5,1,0) & \\
$Q+\overline{Q}$ & & (1/5,3,2)$\eta$ \\
$U+\overline{U}$ & & (8/5,0,1)$\eta$ \\
$D+\overline{D}$ & & (2/5,0,1)$\eta$ \\
$L+\overline{L}$ & & (3/5,1,0)$\eta$ \\
$E+\overline{E}$ & & (6/5,0,0)$\eta$ \\ \hline
total: & (3/5,-3,-6) & (4,4,4)$\eta$  \\ \hline\hline
\end{tabular}
\end{center}
\caption{Contributions to $\tilde{b}_i$.}
\end{table}
This table is useful in that it allows us to mix and match.  We will
do so taking into account the string constraint that bulk matter may 
only transform under bulk gauge groups.  For example,
consider a model with all leptons and gauge fields living in the bulk, but
with Higgs fields and quarks at the fixed points.  The $\tilde{b}_i$
are given by
\begin{equation}
\tilde{b}_i=(0,-4,-6)+3\cdot(9/5,1,0) = (27/5,-1,-6) \,\,\, .
\label{eq:bsc1}
\end{equation}
In this case we find
\begin{equation}
\frac{B_{12}}{B_{13}}=\frac{128}{133}\approx 0.96 \,\,\,\,\, \mbox{ and }
\,\,\,\,\,
\frac{B_{13}}{B_{23}}=\frac{19}{20}\approx 0.95 \,\,\, .
\end{equation}
As we will confirm explicitly in the next section, this scenario 
achieves unification more precisely than the minimal one.

In Table~2, we present three scenarios with $B_{ij}$ ratios that are
significantly better than in the minimal scenario.
\begin{table}
\begin{center}
\begin{tabular}{ccccc}\hline\hline
Scenario & Bulk MSSM Fields & Exotic Fields
& $B_{12}/B_{13}$ & $B_{13}/B_{23}$ \\ \hline
Minimal & SU(3), SU(2) ,U(1), H & none &0.94 & 0.92 \\
1       & SU(3), SU(2), U(1), 3E, 3L & none & 0.96 & 0.95 \\
2       & SU(3), U(1), U, D, 3E & none & 1.00 & 1.00 \\
3       & SU(2), U(1), H, 3L, E & Two $5+\overline{5}$ w/ blk leptons 
& 1.00 & 1.00 \\
\hline\hline
\end{tabular}
\end{center}
\caption{Three Scenarios.  For explanation of the notation, 
see the text.}
\end{table}
We indicate the gauge group when the corresponding gauge multiplet is 
a bulk field, H for both MSSM Higgs fields, and $n\Phi$ for $n$ generations 
of an MSSM matter field $\Phi \equiv (Q,U,D,L, \mbox{ or } E)$. 
Note that it is possible in scenario~1 to exchange an L for an H; the 
vector-like tower of KK excitations associated with a zero-mode left-handed 
lepton field have the same effect on the $\tilde{b}$ as the tower 
associated with the MSSM Higgs fields.  Scenarios~2 and ~3 demonstrate 
that it is not necessary to assume that both SU(2) and SU(3) gauge multiplets
live in the higher dimensional bulk in order to obtain a successful
unification.  As far as we are aware, this point has not been made in the
literature.  Note that only the third scenario involves extra matter,
two SU(5) {\bf 5}+${\bf \overline{5}}$ pairs in which only the leptons
live in the bulk.  We assume that the exotic matter zero modes have a
mass of $\sim m_{{\rm top}}$ for the purpose of our subsequent analysis.  
More strikingly, Scenarios~1 and~2 demonstrate that is possible to achieve an 
improved unification in nonminimal models without the addition of any 
exotic matter multiplets, beyond the mirror fields required to render 
the KK towers of the matter fields vector-like.  We will now consider all 
three scenarios quantitatively, and show that none require large threshold 
corrections at the unification scale.

\section{Numerical Results}

Our numerical analysis of gauge unification in the scenarios listed
in Table~2 is quite conventional.  We adopt the $\overline{MS}$
values for the (GUT normalized) gauge couplings 
$\alpha_1(m_z)=58.99\pm 0.04$ and $\alpha_2(m_z)=29.57\pm 0.03$
that follow from data in the 1998 Review of Particle 
Physics \cite{rpp98}.  We run these up to the top quark mass, where 
we then assume the beta functions of the supersymmetric standard model, and 
where we convert the gauge couplings to the $\overline{DR}$ scheme.  We 
take into account threshold effects, due to varying superparticle masses,
at the one-loop level, and running between $m_{{\rm top}}$ and the
compactification scale $\mu_0$ at the two loop level.  We then use 
Eq.~(\ref{eq:drun}) above the scale $\mu_0$ to determine the
unification point.  Thus, our procedure is similar to Ref.~\cite{ross},
except that we allow for greater freedom in our choice of weak scale
threshold corrections.  This procedure is iterated with trial values
of $\alpha_3(m_z)$ until a suitable three coupling unification is
achieved.  For each of the given scenarios we obtain a prediction
for $\alpha_3(m_z)$ assuming no threshold corrections at the 
unification scale.  While such high scale threshold corrections
should be present generically, our approach allows us to test the
assumption that these need not be large.

\begin{figure}  
\centerline{ \epsfxsize=6.5 in \epsfbox{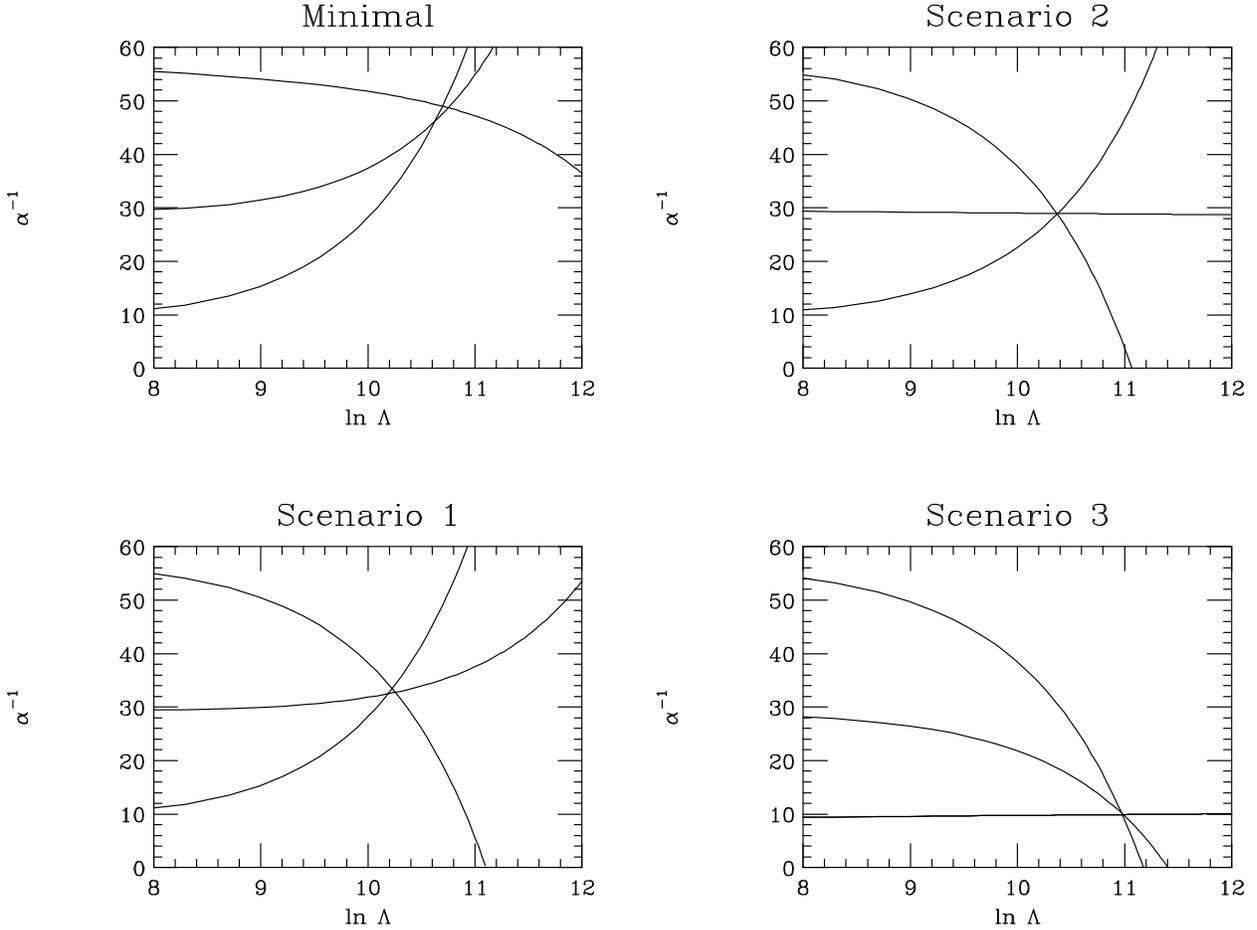}  }
\vglue .3 in
\caption{Unification in Scenarios 1, 2 and 3, with $\Lambda$ in GeV.  The 
minimal scenario is provided for comparison.}

\end{figure}

In Fig.~1, we show the qualitative behavior of unification in 
scenarios 1, 2 and 3 by plotting the running couplings above a
compactification scale of $2$~TeV, assuming the experimental 
value of $\alpha_3(m_z)=0.1191\pm 0.0018$~\cite{rpp98}.  Table~3 presents 
predictions for $\alpha_3(m_z)$ assuming either an intermediate or low 
unification scale.  We display results for $\delta=1$ and $2$, and for 
$\mu_0=2$~TeV and $10^8$~GeV.  These choices are sufficient to understand the
qualitative behavior of the results: as $\delta$ increases, the
predictions for $\alpha_3(m_z)$ increase monotonically, while for
increasing values of $\mu_0$, the predictions approach that of the
MSSM without extra dimensions.  Table~4 provides the predictions for
$\alpha_3(m_z)$ including representative weak scale threshold effects,
in which we've either placed all the non-colored MSSM superpartners 
at $1$~TeV, with the rest at $m_{{\rm top}}$, or vice versa.  Let us consider 
the results for each of the scenarios in turn:

\begin{table}
\begin{center}
\begin{tabular}{c|cccc}\hline\hline
Scenario & $\delta=1$ &  $\delta=2$ &
$\delta=1$ &  $\delta=2$ \\
& $\mu_0= 2$~TeV & $\mu_0= 2$~TeV & $\mu_0= 10^8$~GeV & $\mu_0= 10^8$~GeV \\
\hline
minimal & 0.1734 & 0.1778 & 0.1520 & 0.1552 \\
1  & 0.1438 & 0.1453 & 0.1377 & 0.1388 \\
2  & 0.1159 & 0.1162 & 0.1199 & 0.1199 \\
3  & 0.1155 & 0.1157 & 0.1169 & 0.1169 \\ \hline\hline
\end{tabular}
\end{center}
\caption{Predictions for $\alpha_3(m_z)$, assuming no
weak scale threshold corrections.}
\end{table}

\begin{table}
\begin{center}
\begin{tabular}{c|cccc}\hline\hline
Scenario & $\delta=1$ &  $\delta=2$ &
$\delta=1$ &  $\delta=2$ \\
& $\mu_0= 2$~TeV & $\mu_0= 2$~TeV & $\mu_0= 10^8$~GeV & $\mu_0= 10^8$~GeV \\
\hline
\multicolumn{5}{c}{{\bf A}} \\ \hline
minimal & 0.1473 & 0.1509 & 0.1320 & 0.1344 \\
1 & 0.1263 & 0.1283 & 0.1206 & 0.1215 \\
2 & 0.1047 & 0.1049 & 0.1076 & 0.1077 \\
3 & 0.1044 & 0.1044 & 0.1055 & 0.1056 \\ \hline
\multicolumn{5}{c}{{\bf B}} \\ \hline
minimal & 0.1947 & 0.2006 & 0.1678 & 0.1717 \\
1 & 0.1566 & 0.1606 & 0.1495 & 0.1509 \\
2 & 0.1255 & 0.1258 & 0.1297 & 0.1297 \\
3 & 0.1243 & 0.1246 & 0.1258 & 0.1258 \\
\hline\hline
\end{tabular}
\end{center}
\caption{Predictions for $\alpha_3(m_z)$, assuming (A)
all noncolored MSSM superpartners have $1$~TeV masses while the
rest have masses of $m_{{\rm top}}$, and (B) all colored MSSM superpartners 
have $1$~TeV masses while the rest have masses of $m_{{\rm top}}$.}
\end{table}

$\bullet$ Minimal scenario:  This is the $\eta=0$ scenario of
DDG, which we include as a point of reference. The beta functions for 
this scenario are given in Eq.~(\ref{eq:bmin}).  Note that
for $\delta=1$ and $\mu_0=2$~TeV the low energy value of $\alpha_3(m_z)$ 
is $\sim$30 standard deviations above the experimental central value, 
$0.1191\pm0.0018$ \cite{rpp98}, and improves to $\sim$16 standard 
deviations\footnote{Had we used value of $\alpha_3(m_z)$ in the 1996
Review of Particle Physics, we would obtain results too high by
about $9.8$ standard deviations.  The experimental determination of
$\alpha_3$ has since improved.} if one assumes that colored MSSM
superpartners at the weak threshold are all at $m_{{\rm top}}$ while 
noncolored sparticles are at $\sim$1~TeV. These results agree qualitatively 
with those in Ref.~\cite{ross}, where a different approximation for weak 
scale threshold effects was used.

$\bullet$ Scenario 1: In this scenario, the gauge fields and
leptons live in the bulk, while the Higgs and quarks live
at orbifold fixed points.  The beta functions for this scenario
were given in Eq.~(\ref{eq:bsc1}).  Notice that our previous
observation that this scenario satisfies the relation
$B_{12}/B_{13}=B_{13}/B_{23}=1$ more accurately than the minimal
case does translate into a better predictions for $\alpha_3(m_z)$.
For $\delta=1$ and $\mu_0=2$~TeV, and assuming the same choice for weak
scale threshold corrections applied to the minimal scenario above, we 
find agreement with the experimental value of $\alpha_3(m_z)$ at the 
$4$~standard deviation level.

$\bullet$ Scenario 2:  In this scenario, the SU(2) gauge multiplet
and Higgs fields are confined to the fixed point, while precisely 
one generation of right-handed up and down quarks, and three generations
of right-handed leptons live in the bulk. The KK beta function contributions 
are given by
\begin{equation}
\tilde{b}_i=(28/5,0,-4) \,\,\, .
\end{equation}
This scenario achieves unification much more precisely than the minimal
one.  In the case where $\mu_0=2$~TeV and $\delta=1$, the predicted value
of $\alpha_3(m_z)$ is only 1.8 standard deviations {\em below} the
experimental central value, ignoring all threshold corrections.  
Allowing the superparticle spectrum to vary as in Table~4, we find
this prediction varies between
\begin{equation}
0.105 <\alpha_3(m_z)< 0.126 \,\,\, .
\end{equation}
Thus, unification can be achieved in this model without any string
scale threshold corrections.

$\bullet$ Scenario 3: In this example, the SU(3) gauge multiplet 
is confined to the orbifold fixed point, and hence we are only allowed 
to place leptons and Higgs fields in the bulk.  If we let $\eta_E$ and
$\eta_L$ represent the number of right-handed and left-handed KK lepton 
excitations (including their chiral conjugate partners), then the 
constraint $B_{12}=B_{13}=B_{23}$ implies that $\eta_E=(3\eta_L-13)/2$, which 
has no solution for only three generations of left-handed lepton fields.  
However, there is a simple way to circumvent this problem.  Notice 
that if we add SU(5) {\bf 5}+${\bf \overline{5}}$ pairs in which only the
lepton compontents live in the bulk, then the condition on
$\eta_E$ and $\eta_L$ given above will hold, since differences in 
zero mode beta function pairs will remain unchanged.  A solution may
then be obtained by choosing  $\eta_L=5$ and $\eta_E=1$, which
implies the existence of two such {\bf 5}+${\bf \overline{5}}$ pairs,
when one generation of right-handed and three generations of left-handed 
MSSM lepton fields are assigned to the bulk. The beta functions for this 
scenario are then given by
\begin{equation}
b_i=(43/5,3,-1) \,\,\,\,\,\,\,\,\,\, \tilde{b}_i=(24/5,2,0) \,\,\, .
\end{equation}
Notice in this case that the SU(3) gauge coupling only evolves 
logarithmically, since $\tilde{b}_3=0$. Unification may be achieved at a 
low scale by virtue of the power-law  evolution of $\alpha^{-1}_1$ and 
$\alpha^{-1}_2$, as can be seen from Fig.~1.  This scenario is about as 
successful as scenario~2, predicting $\alpha_3(m_z)$ only $2.0$ standard 
deviations below the experimental central value, ignoring all threshold 
corrections, and assuming that the exotic matter fields have masses 
of $m_{{\rm top}}$.  Allowing the sparticle mass spectrum to vary 
between $m_{{\rm top}}$ and 1~TeV, as in Table~4, we find that the 
scenario~3 prediction for $\alpha_3(m_z)$ varies between
\begin{equation}
0.104 <\alpha_3(m_z)< 0.124 \,\,\, ,
\end{equation}
for $\delta=1$ and $\mu_0=2$~TeV.  Again, unification is
achieved without the need for threshold corrections at the high 
scale.  Although this scenario does indeed involve some new matter
fields, the choice is relatively minimal, and may be completely
natural from the point of view of string theory.

\section{Discussion}

What is interesting about the scenarios we've presented is that 
gauge unification can be achieved in so many different ways. 
Each of our scenarios unifies more precisely than the minimal DDG model, 
and none requires large (or in two cases any) threshold corrections
at the unification scale.   These models illustrate two other
interesting points as well:  (1) One can achieve unification when some
of the standard model gauge groups are confined to a brane. (2) There
are some models that unify more precisely than DDG that {\em do not}
require any additional matter fields with exotic quantum numbers,
beyond the vector-like KK towers of certain MSSM fields that are chosen
to live in the bulk.  Before concluding, we comment briefly on some of 
the other phenomenological implications of these scenarios when
the compactification scale is low.  For a more complete discussion of
the phenomenology of standard model KK excitations in models
with TeV scale compactification~\cite{ant1}, we refer the reader to 
Refs.~\cite{sk,ant}.

In scenarios~1 and~2 the gluon has a tower of KK excitations, which leads
to a significant bound on the compactification scale.   The KK gluon 
excitations are massive color octet vector mesons, with couplings both to 
zero mode gluons and to all the quarks.  Thus, the KK gluons are in every 
way identical to flavor universal colorons, and are subject to the same 
bounds.  Recall that in coloron models, one obtains a massive color
octet from the spontaneous breaking of SU(3)$\times$SU(3) down to
the diagonal color SU(3).  In the case where the two SU(3) gauge couplings
are equal, the coloron couples to quarks exactly like a gluon, or a
KK gluon.  The couplings of colorons or KK gluons to zero mode gluons
are completely determined by SU(3) gauge invariance, and hence are
also the same.  Thus the relevant bound on the lowest KK gluon excitation
is given by  $M_c > 759$~GeV at the 95\% confidence level \cite{dop},
which follows from consideration of the dijet spectrum at the Tevatron.
This constraint places a lower bound on the scale for all the KK excitations
in scenarios~1 and~2.  We can obtain a similar bound on the 
compactification scale in scenario~3 from the production and hadronic decay
of $W$ boson KK excitations, which have standard model couplings to 
the quarks: $M_{W'} > 600$~GeV \cite{rpp98}.  Other direct collider bounds 
on the compactification scale require a more detailed analysis, given the 
nonstandard  $W'$ and $Z'$ couplings in our models.  This issue will be
considered elsewhere \cite{ccip}.  Scenario~1 is particularly interesting 
when one takes into account that interaction vertices involving fields that 
all live in the higher dimensional bulk respect a conservation of KK number.  
(One can think of this as arising from the conservation of KK momentum 
following from  translational invariance in the extra dimensions.)  Hence, in 
scenario~1, the KK excitations of the electroweak gauge fields cannot 
couple to the lepton zero modes, and we obtain both $Z'$ and $W'$ bosons 
with otherwise standard couplings, that are naturally leptophobic!  These
states would likely be within the reach of the LHC for TeV scale 
compactifications.  The other two scenarios present a more complicated 
phenomenology, since some generations of a given MSSM matter field live 
in the bulk while others live on the brane.  It follows that the KK 
excitations of a standard model gauge field would have generation-dependent 
couplings to the zero mode matter fields, and may contribute to a variety of
quark and lepton flavor-changing processes.  Finally it is worth pointing
out that in scenario~3, the fact that the KK $W$ boson excitations can't
couple to zero mode left-handed lepton fields, also leads to a
leptophobic $W'$.  While the purpose of the present work was to
focus on gauge unification in these nonminimal scenarios, a more 
quantitative discussion of the TeV scale phenomenology of the 
scenarios described here will be presented in a separate 
publication~\cite{ccip}.

\begin{center}
{\bf Acknowledgments} 
\end{center}
We are grateful to Keith Dienes for his comments on the manuscript,
and thank Marc Sher and Carl Carlson for useful discussions.
We thank the National Science Foundation for support under grant 
PHY-9800741. 
%\appendix
%\section{Appendix}

\end{document}